\documentstyle[11pt,epsfig]{article}

\newcommand{\lslash}{\rlap{/} l}
\newcommand{\pslash}{\rlap{/} p}

\begin{document}

\title{\bf Light-cone divergence in twist-3 correlation functions}

\author{L.P.~Gamberg$^{a}$, D.S. Hwang$^{b}$, A.~Metz$^{c}$, 
and M.~Schlegel$^{c}$
 \\[0.3cm]
$^{a}${\it Division of Science, Penn State Berks, Reading, PA 19610, USA}
\\[0.2cm]
$^{b}${\it Department of Physics, Sejong University, Seoul 143--747, Korea}
\\[0.2cm]
$^{c}${\it Institut f\"ur Theoretische Physik II,} \\
{\it Ruhr-Universit\"at Bochum, D-44780 Bochum, Germany}}

\date{\today}
\maketitle

\begin{abstract}
\noindent
It is argued that the definition of the twist-3 transverse momentum 
dependent correlation functions must be modified if they contain 
light-like Wilson lines.
In the framework of a simple spectator model of the nucleon we show 
explicitly the presence of a light-cone divergence for a specific 
twist-3 time-reversal odd parton density.
This divergence emerges for all eight twist-3 T-odd correlators 
and appears also in the case of a quark target in perturbative QCD.
The divergence can be removed by using non-light-like Wilson lines. 
Based on our results we argue that currently there exists no 
established factorization formula for transverse momentum dependent 
twist-3 observables in semi-inclusive DIS and related processes.

\end{abstract}

Partonic correlation functions (distributions and fragmentation
functions) that depend on both the longitudinal and
transverse parton momentum ($p_T$-dependent) have received 
considerable attention over the past years.
The most prominent examples are the (naive) time-reversal odd (T-odd) 
Sivers parton density~\cite{sivers_89,sivers_90} and the Collins 
fragmentation function~\cite{collins_92b}, where the status of
the mere existence of T-odd parton distributions was clarified only 
relatively recently~\cite{brodsky_02a,collins_02}.
A great deal of  theoretical and experimental 
effort has been devoted to determining these functions
 by characterizing and measuring single spin asymmetries in semi-inclusive 
deep-inelastic scattering 
(DIS)~\cite{airapetian_04,alexakhin_05,diefenthaler_05} and azimuthal 
asymmetries in $e^+ e^-$-annihilation~\cite{abe_05}
(see also~\cite{vogelsang_05,efremov_06}). 

The first experimental studies on single spin asymmetries in 
semi-inclusive DIS dealt with the observables $A_{UL}$ 
(unpolarized lepton beam and longitudinally polarized 
target)~\cite{airapetian_99,airapetian_01,airapetian_02,airapetian_05}
and $A_{LU}$ (longitudinally polarized lepton beam and unpolarized 
target)~\cite{avakian_03}. 
By contrast they are suppressed like $1/Q$ (with $Q$ 
denoting the virtuality of the exchanged photon) with respect to the 
leading twist asymmetry $A_{UT}$ (transversely polarized target).
Consequently, a parton model description of these asymmetries invokes
($p_T$-dependent) twist-3 correlation 
functions~\cite{levelt_94,mulders_95,boer_97c} (see, e.g., 
Refs.~\cite{oganessian_98,efremov_00,desanctis_00,ma_00,efremov_01,efremov_02,yuan_03b,gamberg_03c,anselmino_05a} 
for related theoretical work on twist-3 single spin asymmetries).
A complete list of $p_T$-dependent twist-3 correlation functions was 
obtained only 
recently~\cite{goeke_03,afanasev_03,metz_04,bacchetta_04a,goeke_05}.

It is non-trivial to establish QCD-factorization for transverse momentum
dependent hard scattering processes in terms of $p_T$-dependent partonic 
correlators.
While the twist-2 case was studied in some detail in the literature
(see, e.g., Refs.~\cite{collins_81b,collins_84,ji_04a,ji_04b,collins_04}), 
for twist-3 observables only a tree-level formalism is available up to 
now~\cite{levelt_94,mulders_95,boer_97c,boer_03a}.
To some extent  the difficulty is intimately related to the 
task of finding a suitable Wilson line that ensures color gauge 
invariance of the $p_T$-dependent correlation functions (see in 
particular the discussion in~\cite{collins_03}).
In this context it was emphasized already in Ref.~\cite{collins_81c}, 
that light-like Wilson lines typically lead to divergences. 
At low-order in $\alpha_{\rm qcd}$, in calculations of twist-2 
$p_T$-dependent correlators with light-like Wilson lines indeed 
certain diagrams show such a light-cone divergence.  
On the other hand, one-loop computations of the T-odd Sivers function
defined with a light-like Wilson line yield well-behaved 
results~\cite{collins_02,ji_02,boer_02,gamberg_03a,bacchetta_03b,goeke_06}.

The main intention of this note is to delineate the presence of a 
light-cone divergence explicit in the case of twist-3 T-odd correlators 
with light-like Wilson lines.
To this end we perform a one-loop calculation of the particular twist-3
T-odd parton distribution $g^{\perp}$~\cite{bacchetta_04a,goeke_05} which 
appears in the description of the beam spin 
asymmetry $A_{LU}$~\cite{bacchetta_04a}.
Our study is based on a simple diquark spectator model of the nucleon.
The light-cone divergence which appears for $g^{\perp}$ also shows up 
for all twist-3 T-odd parton distributions.
Moreover, one arrives at the same conclusion when using a quark target 
in perturbative QCD.
It is therefore evident that twist-3 $p_T$-dependent correlators 
containing light-like Wilson lines are undefined.
The presence of the light-cone divergence also implies that the frequently 
used tree-level formalism of 
Refs.~\cite{levelt_94,mulders_95,boer_97c,boer_03a} for twist-3 observables 
in semi-inclusive DIS and related processes at least needs a modification.
Moreover, the so-called subtraction 
formalism~\cite{collins_99,ji_04a,ji_04b,collins_04} used for an all-order
description of $p_T$-dependent twist-2 processes cannot directly be applied 
in the twist-3 case.
As a consequence, currently there exists no established factorization 
formula for processes like the longitudinal twist-3 single spin 
asymmetries $A_{LU}$ and $A_{UL}$.
\\

Now, we discuss elements of the calculation of $g^{\perp}$.
The starting point is the quark-quark correlation function
\begin{eqnarray} \label{e:corr}
\Phi_{ij}^{(v)}(x,\vec{p}_{T};P,S) 
& = & \int \frac{d\xi^{-} \, d^{2}\vec{\xi}_{T}}{(2\pi)^{3}} \, e^{i p \cdot \xi} 
\langle P,S | \bar{\Psi}_{j}(0) \,
{\cal P} \exp \bigg\{ -i g \int_{0}^{\infty} d\lambda \, 
                        (v \cdot A)(\lambda v) \bigg\} 
\nonumber \\
&  & \hspace{2cm} \mbox{} \times 
{\cal P} \exp \bigg\{ i g \int_{0}^{\infty} d\lambda \, 
                        (v \cdot A)(\lambda v + \xi) \bigg\} \,
\Psi_{i}(\xi) | P,S \rangle \bigg|_{\xi^+ = 0} ,
\end{eqnarray}
where $P$, $S$, and $p$ respectively denote the nucleon momentum, nucleon 
spin,  and quark momentum.
We choose a frame in which the light-cone momentum $P^+$ is large, and
the plus-component of the quark momentum is given by $p^+ = x P^+$. 
The two Wilson lines that appear in the definition~(\ref{e:corr}) 
are determined by the vector $v = (v^+, v^-, \vec{0}_T)$, with the 
proper light-like limit given by $(0, 1, \vec{0}_T)$.
Note that the twist-3 formalism of 
Refs.~\cite{levelt_94,mulders_95,boer_97c,boer_03a} is working with 
light-like lines.
The Wilson lines are of particular importance to enforce color gauge 
invariance in the quark correlation functions as well as to provide 
a mechanism to generate final state interactions and the associated 
complex phases necessary to characterize the T-odd parton 
densities~\cite{brodsky_02a,collins_02}. 

We perform a one-loop calculation in Feynman gauge for which a transverse
gauge link at  light-cone infinity~\cite{ji_02,belitsky_02} is not relevant.
The function $g^{\perp}$ is defined via the correlator in~(\ref{e:corr}) 
by means of~\cite{bacchetta_04a,goeke_05}
\begin{equation} \label{e:defgperp}
2 \varepsilon_{T}^{ij} p_{Tj} \, g^{\perp}(x,\vec{p}_{T}^{\;2}) = 
- P^{+} \, {\mathrm Tr} \Big( \Phi (x,\vec{p}_{T}) \, 
                              \gamma^{i} \gamma_{5} \Big) ,
\end{equation}
where the conventions $\varepsilon^{0123}=1$ and 
$\varepsilon_{T}^{ij} \equiv \varepsilon^{-+ij}$ are used. 
Note that in~(\ref{e:defgperp}) we take an average over the 
nucleon polarization, as the function $g^{\perp}$ is defined for an
unpolarized target.

In the calculation the vector $v$ is taken to be slightly off the 
light-cone $(v^+ \neq 0 \,, \; |v^+| / v^- \ll 1)$ and space-like 
$(v^2 < 0)$. 
For our purpose here one equally well could use a time-like Wilson line.  
The light-cone divergence of $g^{\perp}$ will show up in the light-like
limit $|v^+| / v^-  \to 0$.
Non-light-like Wilson lines as regulators were already used much earlier 
in connection with a proper definition of $p_T$-dependent correlation 
functions~\cite{collins_81c}.
%%%%%%%%%%%%%%
\begin{figure}[t]
\begin{center}
\includegraphics[width=10cm]{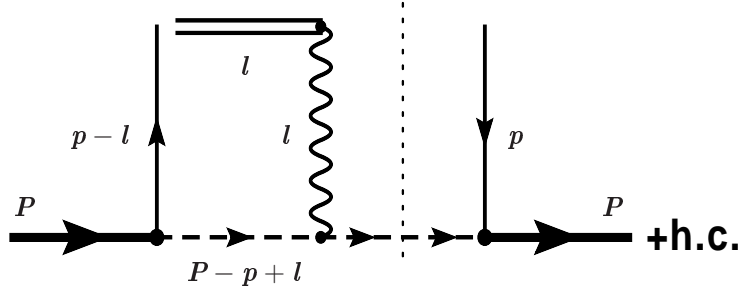}
\end{center}
\caption{One-loop diagram relevant to the calculation of $g^{\perp}$ for the
quark in the scalar diquark model of Ref.~\cite{brodsky_02a}.
[h.c.] stands for the Hermitian Conjugate diagram. 
The thin dashed line depicts the cut of the correlator.
In this diagram the diquark (dashed line) interacts through the exchange of 
a photon with the eikonalized quark (double line).}
\label{f:box}
\end{figure}
%%%%%%%%%%%%%%

We compute the correlator in the simple diquark spectator model of the 
nucleon used in Ref.~\cite{brodsky_02a}.
In this approach the interaction between the nucleon, the quark, and
the diquark is given by a point-like scalar vertex with the coupling
constant $\lambda$.
We use the model with an Abelian gauge boson.
In order to determine the distribution $g^{\perp}$ to lowest non-trivial
order the diagram in Fig.~\ref{f:box} has to be evaluated, where only
the imaginary part of the loop amplitude (with loop-momentum $l$)
is relevant. 
The contribution of this diagram to the correlator is given by
\begin{eqnarray}
\Phi_{ij}^{(v)}(x,\vec{p}_{T};P,S) & = & 
 i \frac{e_q e_s \lambda^2}{2(2\pi)^{3}}
\frac{1}{(\vec{p}_{T}^{\;2} + \tilde{m}^{2}) P^+}
\int \frac{d^{4}l}{(2\pi)^{4}} \,  v \cdot(2P - 2p + l)
\nonumber \\
&  & \mbox{} \times \bigg\{ 
  \frac{ [ \bar{u}(P,S)(\pslash -\lslash + m_q) ]_{j} \,
         [ (\pslash + m_q) u(P,S) ]_{i}}
       { [(l \cdot v) + i \varepsilon] 
         [l^{2} - i \varepsilon]
         [(p - l)^{2} - m_q^{2} - i \varepsilon]
         [(P - p + l)^{2} - m_{s}^{2} - i \varepsilon]}
\nonumber \\
 &  & 
- \frac{ [\bar{u}(P,S)(\pslash + m_q) ]_{j} \,
         [(\pslash - \lslash + m_q) u(P,S) ]_{i}}
       { [(l \cdot v) - i \varepsilon] 
         [l^{2} + i \varepsilon]
         [(p - l)^{2} - m_q^{2} + i \varepsilon]
         [(P - p + l)^{2} - m_{s}^{2} + i \varepsilon]} \bigg\} \,  ,
\end{eqnarray}
with
$\tilde{m}^{2} \equiv x(1-x) \bigg( - M^2 + \frac{m_q^2}{x}                                    + \frac{m_s^2}{1-x} \bigg) $
and quark momentum in light-cone coordinates given by
$p = \bigg( x P^+, \, 
               -\frac{\vec{p}_{T}^{\;2} + m_{s}^{2} - M^2(1-x)}{2P^+(1-x)}, \, 
                \vec{p}_{T} \bigg)$.
Using the definition in Eq.~(\ref{e:defgperp}) the calculation of $g^{\perp}$ 
is straightforward and one arrives at the expression
\begin{eqnarray} \label{e:intgperp}
\lefteqn{ \varepsilon_{T}^{ij} p_{Tj} \, g^{\perp}(x,\vec{p}_{T}^{\;2}) = 
- \frac{e_q e_s \lambda^2}{2(2\pi)^3}
\frac{1}{(\vec{p}_{T}^{\;2} + \tilde{m}^{2})} 
\sum_{\pm} \int \frac{d^{4}l}{(2\pi)^{4}} } 
\nonumber \\
&  & \mbox{} \times
\frac{ v \cdot (2P - 2p + l) 
   \Big[ \varepsilon_{T}^{ij} p_{Tj} (P^{+} l^{-} - P^{-} l^{+} )
        +\varepsilon_{T}^{ij} l_{Tj} (P^{-} p^{+} - P^{+} p^{-} ) \Big] }
     {[(l \cdot v) \pm i \varepsilon]
      [l^{2} \mp i \varepsilon]
      [(p - l)^{2} - m_q^{2} \mp i \varepsilon]
      [(P - p + l)^{2} - m_{s}^{2} \mp i \varepsilon]} \,.
\end{eqnarray} 
The difficulty with the use of light-like lines can be understood
by considering the dependence on the loop integration variable
$l$ in Eq.~(\ref{e:intgperp}).
If the vector $v$ is taken to be light-like the $l^-$-dependence in the 
propagator that is due to the Wilson line drops out.
As a consequence the integral diverges (logarithmically) for $l^+ \to 0$
and $l^- \to \pm \infty$.
The crucial point is the presence of the factor $l^-$ in the numerator 
of~(\ref{e:intgperp}).
This is in contrast to the corresponding calculations of 
the twist-2 T-odd Sivers function $f_{1T}^{\perp}$ and Boer-Mulders
function $h_1^{\perp}$, for which the box graph in Fig.~\ref{f:box} 
provides a well-defined finite result when using light-like Wilson 
lines.
We also note that, as far as a light-cone divergence is concerned, 
the contribution of the box graph to the six twist-2 T-even $p_T$-dependent 
parton densities is well behaved in the light-like limit, because there 
appears no factor of $l^-$ in the numerator.
Here we do not address possible complications (for T-even functions) in
the case of a massless gluon.

Performing the loop-integral in~(\ref{e:intgperp}) yields the result 
\begin{eqnarray} \label{e:resgperp}
g^{\perp}(x,\vec{p}_{T}^{\;2}) & = &
\frac{e_q e_s \lambda^2}{4 (2\pi)^4} 
\frac{1}{(\vec{p}_{T}^{\;2} + \tilde{m}^{2})} 
\bigg\{ \frac{1-x}{x} \, \ln \bigg( \frac{|v^+| \tilde{m}^2}{v^- (P^+)^2} \bigg)
      - \frac{1-x}{x} \bigg( \ln \, (2x (1-x)) 
                 - \frac{2}{2-x} \, \ln \Big( \frac{x}{2}\Big) \bigg)
\nonumber \\
& & 
+ \frac{(1-x) \vec{p}_T^{\;2} + x m_s^2 - x(1-x)^2 M^2}{x \vec{p}_T^{\;2}} 
\ln \bigg( \frac{\vec{p}_T^{\;2} + \tilde{m}^2}{\tilde{m}^2} \bigg) \bigg\}
+ {\cal O} \bigg( \frac{|v^+|}{v^-}\bigg) \,,
\end{eqnarray}
where we made an expansion in the (small) parameter $|v^+| / v^-$ in order 
to simplify the algebra. 
The first term in the brackets in (\ref{e:resgperp}) makes the logarithmic
light-cone divergence explicit.
Note also that the result is boost-invariant and that, due to the expansion 
in $|v^+| / v^-$, the result cannot be used in the limit $x \to 0$.
Eventually, we mention that in the case of the corresponding treatment
of the Sivers function with a non-light-like line we recover 
the well-known (finite) 
result~\cite{collins_02,ji_02,gamberg_03a,bacchetta_03b,goeke_06}
in the light-like limit by putting $|v^+| / v^- \to 0$ at the end of 
the calculation.  

Let us also point out that the light-cone divergence can be avoided by means 
of a phenomenological approach in the sense that we replace the 
proton-quark-diquark coupling according to
\begin{equation} \label{e:monopole}
\lambda \rightarrow \lambda 
 \frac{1}{ \Big( 1 - \frac{p^{2}}{\Lambda^{2}} - i \varepsilon \Big)} \,,
\end{equation}
i.e., we introduce a monopole form factor by hand.
In Eq.~(\ref{e:monopole}), $p$ denotes the quark momentum while $\Lambda$ is an 
arbitrary mass scale. 
Then the expression in~(\ref{e:intgperp}) is replaced by
\begin{eqnarray} \label{e:monogperp}
\lefteqn{ \varepsilon_{T}^{ij} p_{Tj} \, g^{\perp}(x,\vec{p}_{T}^{\;2}) = 
 \frac{e_q e_s \lambda^2}{2 (2\pi)^3}
\frac{\Lambda^4 (1-x)}{(\vec{p}_{T}^{\;2} + \tilde{m}^{2})
         (\vec{p}_{T}^{\;2} + \tilde{m}_{\Lambda}^{2})} 
\sum_{\pm} \int \frac{d^{4}l}{(2\pi)^{4}} } 
\nonumber \\
&  & \mbox{} \times
\frac{ (2 P^+ (1-x) + l^+) 
   \Big[ \varepsilon_{T}^{ij} p_{Tj} (P^{+} l^{-} - P^{-} l^{+} )
        +\varepsilon_{T}^{ij} l_{Tj} (P^{-} p^{+} - P^{+} p^{-} ) \Big] }
     {[l^+ \pm i \varepsilon]
      [l^{2} \mp i \varepsilon]
      [(p - l)^{2} - m_q^{2} \mp i \varepsilon]
      [(P - p + l)^{2} - m_{s}^{2} \mp i \varepsilon]
      [(p - l)^{2} - \Lambda^2 \mp i \varepsilon] } \,,
\\
& & \textrm{with} \quad
\tilde{m}_{\Lambda}^{2} \equiv x(1-x) \bigg( - M^2 + \frac{\Lambda^2}{x} 
                                    + \frac{m_s^2}{1-x} \bigg) \,. 
\nonumber
\end{eqnarray}
In~(\ref{e:monogperp}) we  use the light-like Wilson line
because using this form factor results in an additional factor of $l^-$ 
in the denominator rendering the integral finite in the potentially 
dangerous region $l^+ \to 0$ and $l^- \to \pm \infty$.
Although we get a finite result in this way we want to emphasize again that 
such an approach is purely phenomenological and cannot be derived using a 
Lagrangian of a microscopic theory/model.
\\

We end this short note by summarizing the main points and providing some
additional discussion.
\begin{itemize}
\item We have shown that twist-3 T-odd $p_T$-dependent parton densities 
 cannot be defined by using light-like Wilson lines.
 In order to make this point explicit we have presented a one-loop 
 calculation of the parton distribution $g^{\perp}$, which appears
 for instance in the beam spin asymmetry $A_{LU}$ in semi-inclusive DIS.
 We arrive at the same conclusion for all eight twist-3 T-odd parton 
 densities of a spin-$1/2$ particle for which we pushed the calculation 
 up to the level of Eq.~(\ref{e:intgperp}).
\item While  our calculation  is based on a simple diquark spectator model 
 of the nucleon, we find the same light-cone divergence also in the case of a 
 quark target in perturbative QCD.
 Therefore, the main conclusion we have drawn is not based on some artefact 
 of a simple model.
\item With regard to possible light-cone divergences from Fig.~\ref{f:box}, 
 all eight twist-2 $p_T$-dependent parton densities are well-behaved in the 
 light-like limit.
\item 
 Although we expect that the same divergence problem arises from the graph 
 in Fig.~\ref{f:box} for T-even twist-3 parton densities, at present we 
 cannot exclude that there appears a cancellation of divergent terms once 
 all one-loop diagrams are taken into account.
\item For the T-odd functions we considered here the light-cone divergence 
 can be regularized by using non-light-like Wilson lines 
 (see, e.g., also Ref.~\cite{collins_81c}).
 We introduced such lines by hand.
 It would be very interesting to investigate (as a first step) if a tree-level 
 factorization formalism in the spirit of 
 Refs.~\cite{levelt_94,mulders_95,boer_97c,boer_03a} can be established which 
 automatically provides definitions of twist-3 $p_T$-dependent 
correlators with  non-light-like Wilson lines.
\item In the literature a subtraction method has been used 
 (see, e.g., Refs.~\cite{collins_99,ji_04a,ji_04b,collins_04})
 in order to formulate all-order factorization for transverse momentum 
 dependent twist-2 observables.
 In this formalism also correlation functions with non-light-like lines
 appear.
 However, in particular when performing one-loop calculations of an 
 observable, the dependence on various non-light-like directions drops
 out by construction, as it should be.
 In contrast, using this formalism in the case of, e.g., the longitudinal 
 asymmetry $A_{LU}^{jet}$ for jet production in DIS, which exclusively is 
 given by $g^{\perp}$~\cite{bacchetta_04a}, leads to a result depending on 
 the a priori arbitrary ratio $|v^+|/v^-$.  
 It is quite possible that the subtraction formalism can be generalized
 such that it can cope with transverse momentum dependent twist-3 
 observables.
 But this requires further studies.
\item In a very recent manuscript~\cite{afanasev_06} also a result for
 $g^{\perp}$ in the scalar diquark model has been provided.
 Rather than starting from an operator definition, this result is 
 extracted from the calculation of the beam spin asymmetry $A_{LU}$ in 
 semi-inclusive DIS.
 By doing so and assuming factorization in the spirit of 
 Refs.~\cite{levelt_94,mulders_95,boer_97c,boer_03a} one arrives at an 
 expression for $g^{\perp}$ which does not show the divergence discussed in 
 our note, because the result for the full asymmetry $A_{LU}$ is finite in the 
 diquark spectator model~\cite{afanasev_03,metz_04,afanasev_06}.
 However, apparently the factorization assumption is not valid for this
 twist-3 process, and $g^{\perp}$ obtained in Ref.~\cite{afanasev_06} is not
 the corresponding twist-3 distribution function defined through
 Eqs.~(\ref{e:corr}) and~(\ref{e:defgperp}).

\end{itemize}

\noindent
{\bf Acknowledgements:}
We thank Gary R. Goldstein for useful discussions.
The work of M.S. has been supported by the Graduiertenkolleg
``Physik der Elementarteilchen an Beschleunigern und im Universum.''
The work has also been partially supported by the KOSEF (Korea Science
and Engeneering Foundation), by the Verbundforschung (BMBF),
and by the Transregio/SFB Bochum-Bonn-Giessen.
This research is part of the EU Integrated Infrastructure Initiative
Hadronphysics Project under contract number RII3-CT-2004-506078.

%====== REFERENCES =================================================
%%%%%%%%%%%%%%%%%%%%%%%%%%%%%%%%%%%%%%%%%%%%%%%%%%%%%%%%%%%%%%%%%%%%%%%


\begin{thebibliography}{99}
\bibitem{sivers_89}
D.W.~Sivers, 
Phys. Rev. D {\bf 41}, 83 (1990).

\bibitem{sivers_90}
D.W.~Sivers,
Phys. Rev. D {\bf 43}, 261 (1991).

\bibitem{collins_92b}
J.C.~Collins,
Nucl. Phys. {\bf B396}, 161 (1993). 

\bibitem{brodsky_02a}
S.J.~Brodsky, D.S.~Hwang, and I.~Schmidt,
Phys. Lett. B {\bf 530}, 99 (2002).

\bibitem{collins_02}
J.C.~Collins,
Phys. Lett. B {\bf 536}, 43 (2002).

\bibitem{airapetian_04}
A.~Airapetian {\it et al.}  [HERMES Collaboration],
Phys. Rev. Lett. {\bf 94}, 012002 (2005). 

\bibitem{alexakhin_05}
V.Y.~Alexakhin {\it et al.}  [COMPASS Collaboration],
Phys. Rev. Lett. {\bf 94}, 202002 (2005).

\bibitem{diefenthaler_05}
M.~Diefenthaler,
arXiv:hep-ex/0507013.

\bibitem{abe_05}
R.~Seidl {\it et al.}  [Belle Collaboration],
Phys. Rev. Lett. {\bf 96}, 232002 (2006). 

\bibitem{vogelsang_05}
W.~Vogelsang and F.~Yuan,
Phys. Rev. D {\bf 72}, 054028 (2005).

\bibitem{efremov_06}
A.V.~Efremov, K.~Goeke, and P.~Schweitzer,
Phys. Rev. D {\bf 73}, 094025 (2006).

\bibitem{airapetian_99}
A.~Airapetian {\it et al.}  [HERMES Collaboration],
Phys. Rev. Lett. {\bf 84}, 4047 (2000).

\bibitem{airapetian_01}
A.~Airapetian {\it et al.}  [HERMES Collaboration],
Phys. Rev. D {\bf 64}, 097101 (2001).

\bibitem{airapetian_02}
A.~Airapetian {\it et al.}  [HERMES Collaboration],
Phys. Lett. B {\bf 562}, 182 (2003).

\bibitem{airapetian_05}
A.~Airapetian {\it et al.}  [HERMES Collaboration],
Phys. Lett. B {\bf 622}, 14 (2005).

\bibitem{avakian_03}
H.~Avakian {\it et al.}  [CLAS Collaboration],
Phys. Rev. D {\bf 69}, 112004 (2004).

\bibitem{levelt_94}
J.~Levelt and P.J.~Mulders,
Phys. Lett. B {\bf 338}, 357 (1994). 

\bibitem{mulders_95}
P.J.~Mulders and R.D.~Tangerman,
Nucl. Phys. {\bf B461}, 197 (1996) 
[Erratum-ibid. {\bf B484}, 538 (1997)].

\bibitem{boer_97c}
D.~Boer and P.J.~Mulders,
Phys. Rev. D {\bf 57}, 5780 (1998).

\bibitem{oganessian_98}
K.A.~Oganessian, H.R.~Avakian, N.~Bianchi, and A.M.~Kotzinian,
arXiv:hep-ph/9808368.

\bibitem{efremov_00}
A.V.~Efremov, K.~Goeke, M.V.~Polyakov, and D.~Urbano,
Phys. Lett. B {\bf 478}, 94 (2000).

\bibitem{desanctis_00}
E.~De Sanctis, W.D.~Nowak, and K.A.~Oganessian,
Phys. Lett. B {\bf 483}, 69 (2000).

\bibitem{ma_00}
B.Q.~Ma, I.~Schmidt, and J.J.~Yang,
Phys. Rev. D {\bf 63}, 037501 (2001).

\bibitem{efremov_01}
A.V.~Efremov, K.~Goeke, and P.~Schweitzer,
Phys. Lett. B {\bf 522}, 37 (2001) 
[Erratum-ibid. B {\bf 544}, 389 (2002)].

\bibitem{efremov_02}
A.V.~Efremov, K.~Goeke, and P.~Schweitzer,
Phys. Rev. D {\bf 67}, 114014 (2003).

\bibitem{yuan_03b}
F.~Yuan,
Phys. Lett. B {\bf 589}, 28 (2004). 

\bibitem{gamberg_03c}
L.P.~Gamberg, D.S.~Hwang, and K.A.~Oganessyan,
Phys. Lett. B {\bf 584}, 276 (2004). 

\bibitem{anselmino_05a}
M.~Anselmino, M.~Boglione, U.~D'Alesio, A.~Kotzinian, F.~Murgia,
and A.~Prokudin,
Phys. Rev. D {\bf 71}, 074006 (2005).

\bibitem{goeke_03}
K.~Goeke, A.~Metz, P.V.~Pobylitsa, and M.V.~Polyakov,
Phys. Lett. B {\bf 567}, 27 (2003).

\bibitem{afanasev_03}
A.~Afanasev and C.E.~Carlson,
arXiv:hep-ph/0308163.

\bibitem{metz_04}
A.~Metz and M.~Schlegel,
Eur. Phys. J. {\bf A22}, 489 (2004).

\bibitem{bacchetta_04a}
A.~Bacchetta, P.J.~Mulders, and F.~Pijlman,
Phys. Lett. B {\bf 595}, 309 (2004). 

\bibitem{goeke_05}
K.~Goeke, A.~Metz, and M.~Schlegel,
Phys. Lett. B {\bf 618}, 90 (2005). 

\bibitem{collins_81b}
J.C.~Collins and D.E.~Soper,
Nucl. Phys. {\bf B193}, 381 (1981) 
[Erratum-ibid. {\bf B213}, 545 (1983)].

\bibitem{collins_84}
J.C.~Collins, D.E.~Soper, and G.~Sterman,
Nucl. Phys. {\bf B250}, 199 (1985).

\bibitem{ji_04a}
X.~Ji, J.P.~Ma, and F.~Yuan,
Phys. Rev. D {\bf 71}, 034005 (2005).

\bibitem{ji_04b}
X.~Ji, J.P.~Ma, and F.~Yuan,
Phys. Lett. B {\bf 597}, 299 (2004).

\bibitem{collins_04}
J.C.~Collins and A.~Metz,
Phys. Rev. Lett. {\bf 93}, 252001 (2004).

\bibitem{boer_03a}
D.~Boer, P.J.~Mulders, and F.~Pijlman,
Nucl. Phys. {\bf B667}, 201 (2003).

\bibitem{collins_03}
J.C.~Collins,
Acta Phys. Polon. B {\bf 34}, 3103 (2003).

\bibitem{collins_81c}
J.C.~Collins and D.E.~Soper,
Nucl. Phys. {\bf B194}, 445 (1982).

\bibitem{ji_02}
X.~Ji and F.~Yuan,
Phys. Lett. B {\bf 543}, 66 (2002). 

\bibitem{boer_02}
D.~Boer, S.J.~Brodsky, D.S.~Hwang,
Phys. Rev. D {\bf 67}, 054003 (2003).

\bibitem{gamberg_03a}
L.P.~Gamberg, G.R.~Goldstein, and K.A.~Oganessyan,
Phys. Rev. D {\bf 67}, 071504 (2003).

\bibitem{bacchetta_03b}
A.~Bacchetta, A.~Schaefer, and J.J.~Yang,
Phys. Lett. B {\bf 578}, 109 (2004).

\bibitem{goeke_06}
K.~Goeke, S.~Meissner, A.~Metz, and M.~Schlegel,
Phys. Lett. B {\bf 637}, 241 (2006).

\bibitem{belitsky_02}
A.V.~Belitsky, X.~Ji, and F.~Yuan,
Nucl. Phys. {\bf B656}, 165 (2003).

\bibitem{collins_99}
J.C.~Collins and F.~Hautmann,
Phys. Lett. B {\bf 472}, 129 (2000). 

\bibitem{afanasev_06}
A.V.~Afanasev and C.E.~Carlson,
arXiv:hep-ph/0603269.

\end{thebibliography}
\end{document}